\documentclass[12pt]{article}

\usepackage{amssymb}
\usepackage{amsmath}
\usepackage{epsfig}

\begin{document}

\title{Second Order General Slow-Roll Power Spectrum}

\author{
Jeongyeol Choe,
Jinn-Ouk Gong\footnote{jgong@muon.kaist.ac.kr} \ and
Ewan D. Stewart\\
{\em Department of Physics, KAIST, Daejeon 305-701, Republic of
Korea} }

\date{\today}

\maketitle

\begin{abstract}
Recent combined results from the Wilkinson Microwave Anisotropy
Probe (WMAP) and Sloan Digital Sky Survey (SDSS) provide a
remarkable set of data which requires more accurate and general
investigation. Here we derive formulae for the power spectrum
$\mathcal P(k)$ of the density perturbations produced during
inflation in the general slow-roll approximation with second order
corrections. Also, using the result, we derive the power spectrum
in the standard slow-roll picture with previously unknown third
order corrections.
\end{abstract}

\vspace*{-65ex}
\hspace*{\fill} KAIST-TH/2004-05

\thispagestyle{empty}
\setcounter{page}{0}
\newpage
\setcounter{page}{1}

\section{Introduction}

It is generally believed that inflation \cite{inf} provides the
explanation for many of the unexplained features of the standard
hot big bang cosmology. Inflation generates the flatness,
homogeneity and isotropy of the largest observable scales today.
Furthermore, primordial quantum fluctuations, which were stretched
during inflation, are thought to be the origin of the small
perturbations necessary for the formation of galaxies, clusters,
and other rich structures in the universe. The power spectrum of
these perturbations, as detected in many cosmic microwave
background observations and galaxy surveys \cite{cmb}, is observed
to be approximately scale invariant.

The first year data from NASA's Wilkinson Microwave Anisotropy
Probe (WMAP) \cite{wmap} and the Sloan Digital Sky Survey (SDSS)
\cite{sdss} are constraining the power spectrum and the spectral
index with greater accuracy than ever before. Recent estimates
from several observations are presented in \cite{estimate}.
Therefore it is
necessary to obtain more precise and general estimation for the
power spectrum to use the WMAP results fully and prepare for the
future study of the power spectrum.

There has been extensive work on the power spectrum for the
density perturbations produced during inflation
\cite{ps,paction,ps1st,ps3,ps2nds}, using the standard slow-roll
picture. The standard slow-roll approximation assumes the
slow-roll parameters,
\begin{equation}
\epsilon \equiv -\frac{\dot H}{H^2} = -\frac{d \ln H}{d \ln a}
\hspace{1cm} \mbox{and} \hspace{1cm} \delta_1 \equiv
\frac{\ddot\phi}{H\dot\phi} = \frac{d \ln\dot\phi}{d \ln a},
\end{equation}
are small. It also makes the {\em extra\/} assumption that
$\delta_1$ is approximately constant. In the general slow-roll
scheme \cite{gsr}, we abandon this extra assumption so that we can
consider a wider class of inflation models. In this paper, we use
the basic formalism presented in \cite{ps2nds,gsr,ps2ndm} and
extend previous results \cite{gsr} to calculate the power spectrum
to second order in the general slow-roll expansion.

We set $c = \hbar = 8\pi G = 1$ throughout this paper.

\section{General slow-roll formulae for the spectrum}

In this section, we will derive the formula for the power spectrum
of the density perturbations. We briefly review how to proceed
from the fundamental equation using the formalism of
Refs.~\cite{ps2nds,gsr,ps2ndm}. Then, we present our integral
equation for the power spectrum in the context of the general
slow-roll expansion \cite{gsr}.

We begin with the effective action for the inflaton field $\phi$
during inflation,
\begin{equation}
S = \int \left[ -\frac{1}{2}R + \frac{1}{2}(\partial_\mu \phi)^2 -
V(\phi) \right] \sqrt{-g} \ d^4 x,
\end{equation}
from which we can derive the action for the scalar perturbations
as \cite{paction}
\begin{equation}\label{spa}
S = \int \frac{1}{2} \left[ \left(
\frac{\partial\varphi}{\partial\eta} \right)^2 - \left(
\nabla\varphi \right)^2 + \left( \frac{1}{z} \frac{d^2z}{d\eta^2}
\right)\varphi^2 \right] d\eta\,dx^3,
\end{equation}
where
\begin{equation}
\varphi = a \left( \delta\phi + \frac{\dot\phi}{H} \mathcal{R}
\right),
\end{equation}
$\mathcal{R}$ is the intrinsic curvature perturbation of the
spatial hypersurfaces, and
\begin{equation}
z = \frac{a\dot\phi}{H}
\end{equation}
Hence the intrinsic curvature perturbation of the comoving
hypersurfaces is given by
\begin{equation}
\mathcal{R}_\mathrm{c} = \frac{\varphi}{z}.
\end{equation}

Eq.~(\ref{spa}) gives the equation of motion for the Fourier
component of $\varphi$
\begin{equation}\label{eom}
\frac{d^2\varphi_k}{d\eta^2} + \left( k^2 - \frac{1}{z} \frac{d^2
z}{d\eta^2} \right) \varphi_k = 0,
\end{equation}
where $ \varphi_k $ satisfies the boundary conditions
\begin{equation}
\varphi_k \longrightarrow
\left\{ \begin{array}{ccc}
\frac{1}{\sqrt{2k}\,} e^{-ik\eta} & \mbox{as} & -k\eta \rightarrow
\infty
\\
A_k z & \mbox{as} & -k\eta \rightarrow 0.
\end{array}\right.
\end{equation}
Defining $ y \equiv \sqrt{2k} \, \varphi_k $, $ x \equiv -k \eta$
and
\begin{equation}\label{f}
f(\ln x) = \frac{2\pi x z}{k} = \frac{2\pi}{H} \frac{ax\dot\phi}{k},
\end{equation}
Eq.~(\ref{eom}) becomes
\begin{equation}\label{eom2}
\frac{d^2y}{dx^2} + \left( 1 - \frac{2}{x^2} \right)y =
\frac{g(x)}{x^2} y,
\end{equation}
where
\begin{equation}\label{g}
g = \frac{f'' - 3f'}{f}
\end{equation}
and $f' \equiv df/d\ln x$. Using Green's method, we can present
the solution of Eq.~(\ref{eom2}) as an integral equation
\begin{eqnarray}\label{sol}
y(x) & = & y_0(x) + \frac{i}{2} \int_{x}^{\infty} \frac{du}{u^2}
\, g(u) \left[ y_0^*(u) \, y_0(x) - y_0^*(x) \, y_0(u) \right]
y(u)
\\ & \equiv & y_0(x) + L(x,u) \, y(u),
\end{eqnarray}
where
\begin{equation}\label{hsol}
y_0(x) = \left(1 + \frac{i}{x}\right) e^{ix}
\end{equation}
is the homogeneous solution with correct asymptotic behaviour.

The power spectrum for the curvature perturbation $\mathcal{P}(k)$
is defined by
\begin{equation}
\frac{2\pi^2}{k^3}\,\mathcal{P}(k)\,\delta^{(3)}(\mathbf{k-l}) =
\langle \mathcal{R}_\mathrm{c}(\mathbf{k})\,
{\mathcal{R}_\mathrm{c}}^\dagger(\mathbf{l}) \rangle \,,
\end{equation}
which, using the above results, we can write conveniently as
\begin{equation}\label{psy}
\mathcal{P}(k) = \lim_{x \rightarrow 0} \left| \frac{xy}{f}
\right|^2.
\end{equation}

We assume that $y(x)$ is given approximately by the scale
invariant $y_0(x)$, or equivalently that $g$ is small. Then, since
we are interested in the second order corrections, we iterate
Eq.~(\ref{sol}) twice, i.e.
\begin{equation}\label{yx2}
y(x) \simeq y_0(x) + L(x,u) \, y_0(u) + L(x,u) \, L(u,v) \, y_0(v).
\end{equation}
Substituting into Eq.~(\ref{psy}), and using the method of
Ref.~\cite{gsr}, i.e. expanding in powers of $x$ and using the
limit $x \rightarrow 0$, we get
\begin{eqnarray}\label{PSg-before1}
\lefteqn{ \ln \mathcal{P}(k) = \lim_{x \rightarrow 0} \left\{ \ln
\left( \frac{1}{f^2} \right) + \frac{2}{3} \frac{f'}{f} +
\frac{1}{9} \left( \frac{f'}{f} \right)^2 + \frac{2}{3}
\int_x^\infty \frac{du}{u} \, W(u) \, g(u) + \frac{2}{9} \left[
\int_x^\infty \frac{du}{u} \, X(u) \, g(u) \right]^2 \right. } \nonumber
\\ & &
\hspace{2cm} \left. \mbox{} - \frac{2}{3} \int_x^\infty
\frac{du}{u} \, X(u) \, g(u) \int_u^\infty \frac{dv}{v^2} \, g(v) -
\frac{2}{3} \int_x^\infty \frac{du}{u} \, X(u) \, g(u) \int_u^\infty
\frac{dv}{v^4} \, g(v) + \mathcal{O}(g^3) \right\}, \nonumber
\\
\end{eqnarray}
where
\begin{equation}
W(x) = \frac{3\sin(2x)}{2x^3} - \frac{3\cos(2x)}{x^2} -
\frac{3\sin(2x)}{2x}
\end{equation}
and
\begin{equation}
X(x) = - \frac{3\cos(2x)}{2x^3} - \frac{3\sin(2x)}{x^2} +
\frac{3\cos(2x)}{2x} + \frac{3}{2x^3} \left(1+x^2\right).
\end{equation}
Note that
\begin{equation}
\lim_{x \rightarrow 0} W(x) = 1 + \mathcal{O}\left(x^2\right)
\hspace{1cm} \mbox{and} \hspace{1cm} \lim_{x \rightarrow 0} X(x) =
\frac{1}{3} x^3 + \mathcal{O}\left(x^5\right).
\end{equation}
The right hand side of Eq.~(\ref{PSg-before1}) as a whole is well
defined in the limit $x \rightarrow 0$, but we do not know how $f$
behaves in that limit, so the individual terms are not well
defined. As a remedy, we pick some reasonable point, e.g. around
horizon crossing, to evaluate $f$ and rearrange to make the
individual terms well defined. Using
\begin{eqnarray}
\frac{1}{f^2} & = & \frac{1}{f_\star^2} \exp \left[ 2 \ln \left(
\frac{f_\star}{f} \right) \right] \nonumber
\\
& = & \frac{1}{f_\star^2} \left\{ 1 + 2 \int_x^{x_\star}
\frac{du}{u} \frac{f'}{f} + 2 \left( \int_x^{x_\star} \frac{du}{u}
\frac{f'}{f} \right)^2 + \mathcal{O} \left[ \left( \frac{f'}{f}
\right)^3 \right] \right\}
\end{eqnarray}
and
\begin{equation}
\frac{f'}{f} = \frac{f'_\star}{f_\star} - \int_x^{x_\star}
\frac{du}{u} \frac{f''}{f} + \int_x^{x_\star} \frac{du}{u} \left(
\frac{f'}{f} \right)^2,
\end{equation}
where subscript $\star$ denotes evaluation at some convenient time
around horizon crossing, we can rewrite the power spectrum as
\begin{eqnarray}\label{PSg1}
\ln \mathcal{P}(k) & = & \ln \left( \frac{1}{f_\star^2} \right) +
\frac{2}{3} \frac{f'_\star}{f_\star} + \frac{1}{9} \left(
\frac{f'_\star}{f_\star} \right)^2 + \frac{2}{3} \int_0^\infty
\frac{du}{u} \, W_\theta(x_\star,u) \, g(u) + \frac{2}{9} \left[
\int_0^\infty \frac{du}{u} \, X(u) \, g(u) \right]^2 \nonumber
\\ & &
\mbox{} - \frac{2}{3}\int_0^\infty \frac{du}{u} \, X(u) \, g(u)
\int_u^\infty \frac{dv}{v^2} \, g(v) - \frac{2}{3} \int_0^\infty
\frac{du}{u} \, X_\theta(x_\star,u) \, g(u) \int_u^\infty
\frac{dv}{v^4} \, g(v),
\end{eqnarray}
where
\begin{equation}
W_\theta(x_\star,x) \equiv W(x) - \theta(x_\star-x)
\end{equation}
and
\begin{equation}
X_\theta(x_\star,x) \equiv X(x) - \frac{x^3}{3} \, \theta(x_\star-x).
\end{equation}

Eq.~(\ref{PSg1}) can be written more compactly by substituting
\begin{equation}
g = \left( \frac{f'}{f} \right)' - 3 \frac{f'}{f} + \left(
\frac{f'}{f} \right)^2,
\end{equation}
integrating by parts, and using the identity
\begin{equation}\label{iden}
2 x^3 W = \left( 3 + x^2 \right) X + x \left( 1 + x^2 \right) X',
\end{equation}
to give
\begin{eqnarray}\label{simplest}
\ln \mathcal{P}(k) & = & \ln \left(\frac{1}{f_\star^2}\right) - 2
\int_0^\infty \frac{du}{u} \, w_\theta(x_\star,u) \, \frac{f'}{f}
+ 2 \left[ \int_0^\infty \frac{du}{u} \, \chi(u) \, \frac{f'}{f}
\right]^2 \nonumber
\\ & &
\mbox{} - 4 \int_0^\infty \frac{du}{u} \, \chi(u) \, \frac{f'}{f}
\int_u^\infty \frac{dv}{v^2} \, \frac{f'}{f}
\\
& = & - \int_0^\infty du \, W'(u) \left[ \ln \left( \frac{1}{f^2}
\right) + \frac{2}{3} \frac{f'}{f} \right] + 2 \left[
\int_0^\infty \frac{du}{u} \, \chi(u) \, \frac{f'}{f} \right]^2
\nonumber
\\ & &
\mbox{} - 4 \int_0^\infty \frac{du}{u} \, \chi(u) \, \frac{f'}{f}
\int_u^\infty \frac{dv}{v^2} \, \frac{f'}{f}
\end{eqnarray}
where
\begin{equation}
w(x) \equiv W(x) + \frac{x}{3}\,W'(x) = \frac{\sin(2x)}{x} - \cos(2x),
\end{equation}
and
\begin{equation}
\chi(x) \equiv X(x) + \frac{x}{3} X'(x) =
\frac{1}{x} - \frac{\cos(2x)}{x} - \sin(2x).
\end{equation}
These functions behave asymptotically as
\begin{equation}
\lim_{x \rightarrow 0} w(x) = 1 + \mathcal{O}\left(x^2\right)
\hspace{1cm} \mbox{and} \hspace{1cm} \lim_{x \rightarrow 0}
\chi(x) = \frac{2}{3} x^3 + \mathcal{O}\left(x^5\right),
\end{equation}
and we define
\begin{equation}
w_\theta(x_\star,x) \equiv w(x) - \theta(x_\star-x).
\end{equation}

Eqs.~(\ref{PSg1}) and~(\ref{simplest}) are our main results. To
evaluate these formulae, one needs to determine $f$ and $g$ as
functions of $x$. This can be done by solving the background
equations for $a$ and $\phi$ as functions of $x$, and using
Eqs.~(\ref{f}) and~(\ref{g}).

\subsection{Special case of a de Sitter background}

In the physically important special case of constant $H$, we have
\begin{eqnarray}
\frac{1}{f^2} & = & \left(\frac{H}{2\pi}\right)^2
\left(\frac{H}{\dot\phi}\right)^2 \, , \nonumber
\\
\frac{f'}{f} & = & - \frac{\ddot\phi}{H\dot\phi}\, , \nonumber
\\
g & = & - 3 \frac{V''}{V} \, ,
\end{eqnarray}
and
\begin{equation}
\ln \left( \frac{V^3}{12\pi^2V'^2} \right) \simeq \ln \left(
\frac{1}{f^2} \right) + \frac{2}{3} \frac{f'}{f} + \frac{1}{9}
\left( \frac{f'}{f} \right)^2.
\end{equation}
Then the power spectrum Eq.~(\ref{PSg1}) becomes
\begin{eqnarray}\label{pspot}
\ln \mathcal{P} & = & \ln \left(
\frac{V_\star^3}{12\pi^2{V'_\star}^2} \right) - 2 \int_0^\infty
\frac{du}{u} \, W_\theta(x_\star,u) \frac{V''}{V} + 2 \left[
\int_0^\infty \frac{du}{u} \, X(u) \, \frac{V''}{V} \right]^2
\nonumber
\\ \label{PV} & &
\mbox{} - 6 \int_0^\infty \frac{du}{u} \, X(u) \frac{V''}{V}
\int_u^\infty \frac{dv}{v^2} \, \frac{V''}{V} - 6 \int_0^\infty
\frac{du}{u} \, X_\theta(x_\star,u) \, \frac{V''}{V} \int_u^\infty
\frac{dv}{v^4} \frac{V''}{V} \, ,
\end{eqnarray}
while Eq.~(\ref{simplest}) becomes
\begin{eqnarray}
\ln \mathcal{P} & = & \ln \left[ \left( \frac{H}{2\pi} \right)^2
\left( \frac{H}{\dot\phi_\star} \right)^2 \right] + 2
\int_0^\infty \frac{du}{u} w_\theta (x_\star, u)
\frac{\ddot\phi}{H\dot\phi} + 2 \left[ \int_0^\infty \frac{du}{u}
\chi(u) \frac{\ddot\phi}{H\dot\phi} \right]^2 \nonumber
\\ & &
- 4 \int_0^\infty \frac{du}{u} \chi(u) \frac{\ddot\phi}{H\dot\phi}
\int_u^\infty \frac{dv}{v^2} \frac{\ddot\phi}{H\dot\phi} \ .
\end{eqnarray}

\section{Example}

\begin{figure}
\begin{center}
\epsfig{file=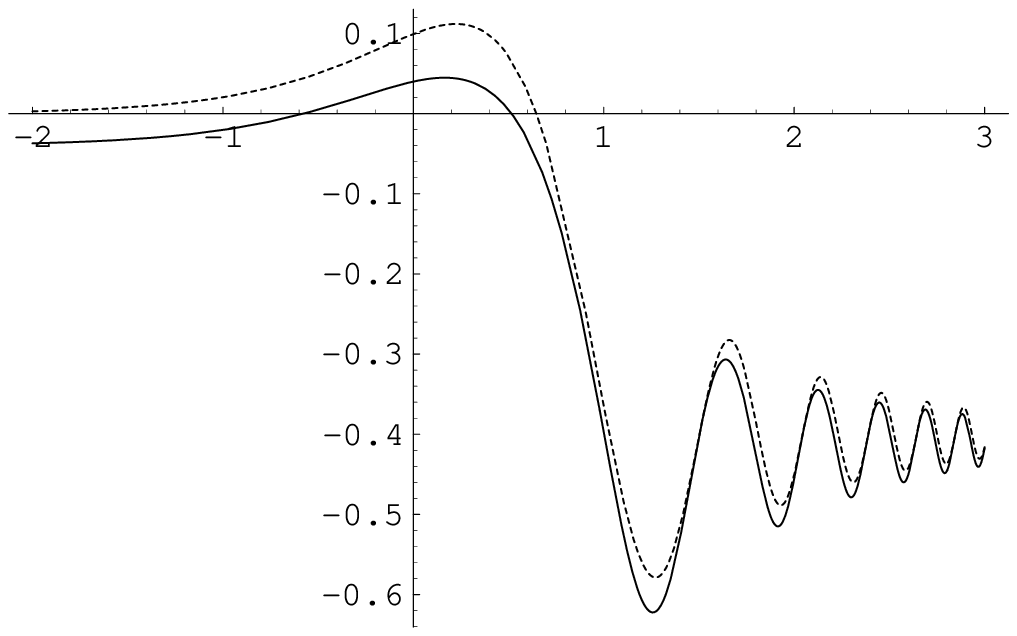, width = 8cm}
\epsfig{file=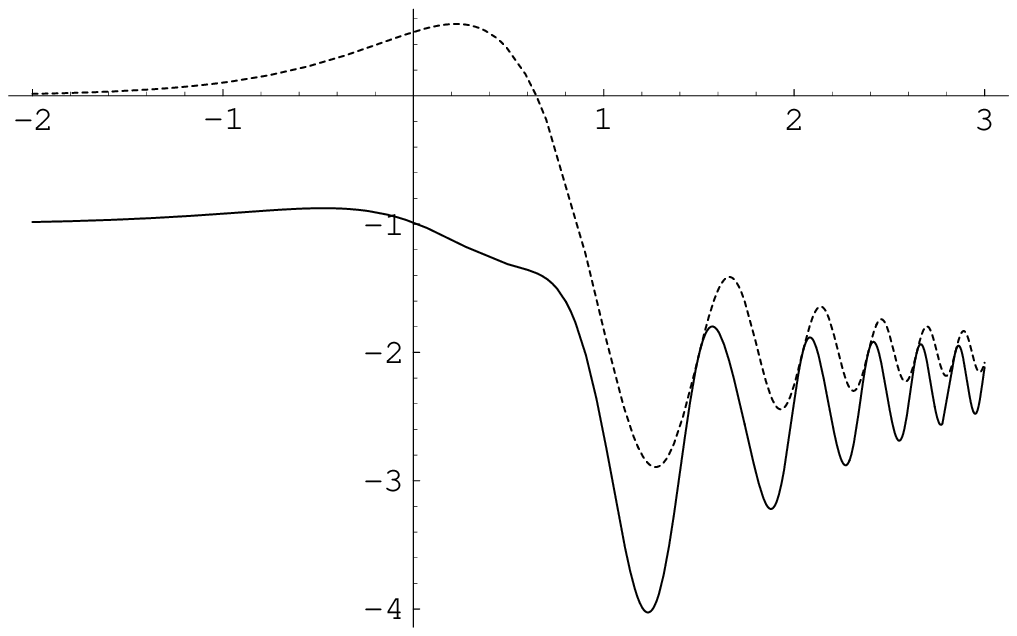, width = 8cm}
\end{center}
\begin{center}
\epsfig{file=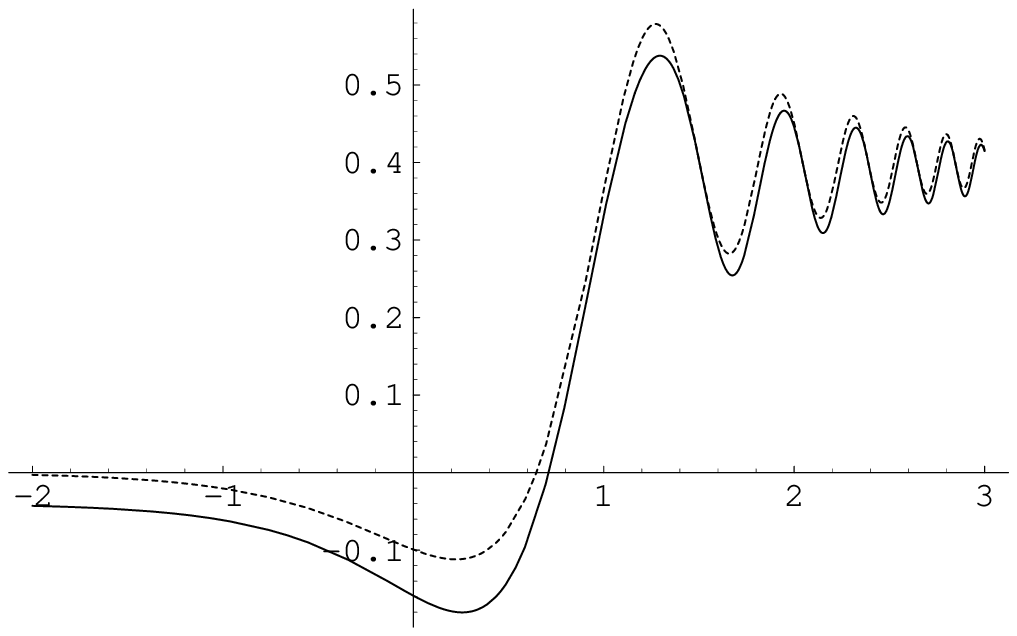, width = 8cm}
\epsfig{file=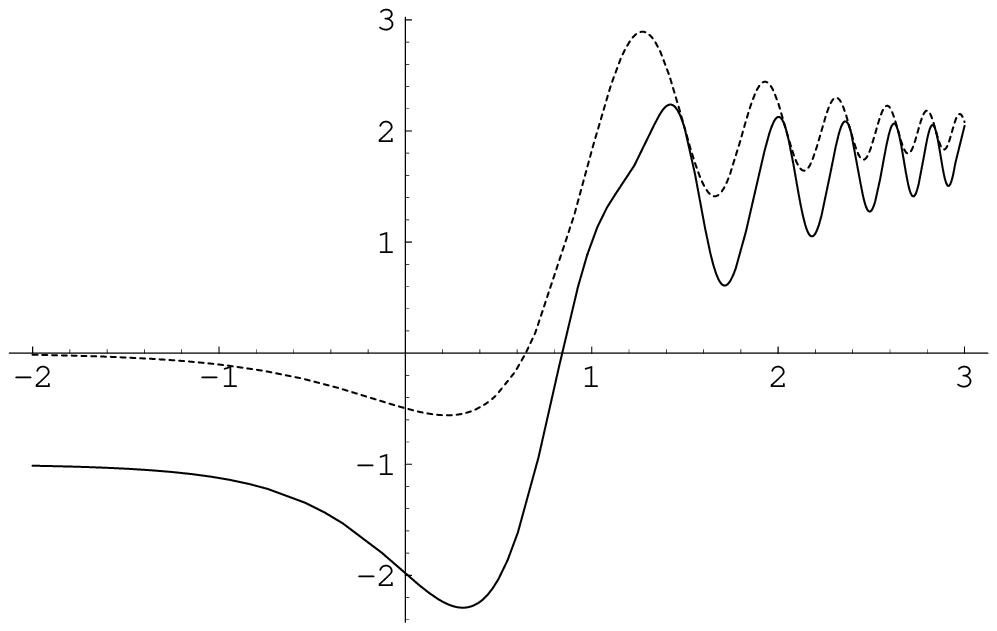, width = 8cm}
\end{center}
\caption{\label{starobinsky} Plot of $\ln \mathcal{P}$ versus $\ln
k$. The solid line is the full result, Eq.~(\ref{ex1ln}), and the
dotted line is the first order term in $\Delta A / A$ only.
$\Delta A / A > 0$ in the upper row and $\Delta A / A < 0$ in the
lower row. $|\Delta A / A| = 0.2$ in the left column and $|\Delta
A / A| = 1$ in the right column.}
\end{figure}

As a specific example, we consider an inflaton $\phi$ rolling down
a linear potential \cite{str} with slope changing from $-A$ to $-
A - \Delta A$ at $\phi = \phi_0$. The potential is
\begin{equation}
V(\phi) = V_0 \left\{ 1 - \left[ A + \theta(\phi-\phi_0) \, \Delta
A \right] \left( \phi - \phi_0 \right) \right\}.
\end{equation}
Assuming $|A| \ll 1$ so that $H \simeq \sqrt{V_0/3}$, and solving
the equation of motion for $\phi$, we obtain
\begin{equation}
\frac{d\phi}{dN} = A + \theta(N-N_0) \, \Delta A \left[ 1 -
e^{-3(N-N_0)} \right],
\end{equation}
where $N = \int H \, dt$ and $\phi(N_0) = \phi_0$. Then if
$|\Delta A / A| \ll 1$, so that the approximate scale invariance
of the spectrum is maintained, we have
\begin{equation}\label{v''}
\frac{V''}{V} = - \Delta A \, \delta(\phi-\phi_0)
= - \frac{\Delta A}{A} \, \delta(N-N_0).
\end{equation}
Substituting into Eq.~(\ref{pspot}) and performing the
integration, we find the power spectrum
\begin{eqnarray}\label{ex1ln}
\ln \mathcal{P} = \ln \left[ \frac{V_0}{12\pi^2(A+\Delta A)^2}
\right] + 2\left( \frac{\Delta A}{A} \right) W(x_0) + 2 \left(
\frac{\Delta A}{A} \right)^2 X(x_0) \left[ X(x_0) -
\frac{3}{2x_0^3} ( 1 + x_0^2 ) \right], \nonumber
\\
\end{eqnarray}
where the $x_\star$ dependent terms in Eq.~(\ref{pspot}) have been
absorbed into the first term, the constant leading result, to show
explicitly that the spectrum is independent of the evaluation
point $x_\star$. The spectrum $\ln\mathcal{P}$ is plotted in
Figure~\ref{starobinsky}.

Using the identity
\begin{equation}
W^2 + X^2 = \frac{3}{x^3} \left( 1 + x^2 \right) X,
\end{equation}
we can write Eq.~(\ref{ex1ln}) as
\begin{equation}
\mathcal{P} = \frac{V_0}{12\pi^2(A + \Delta A)^2} \left[ 1 + 2
\left( \frac{\Delta A}{A} \right) W(x_0) + 3\left( \frac{\Delta
A}{A} \right)^2 \frac{1 + x_0^2}{x_0^3} \, X(x_0) \right],
\end{equation}
which is exactly the same as found in Ref.~\cite{str}.

\section{Application to the standard slow-roll expansion}

In this section, we apply our general slow-roll formula to the
special case of standard slow-roll. Our general slow-roll formula,
Eq.~(\ref{simplest}), can be written as
\begin{eqnarray}\label{PSf}
\mathcal{P} & = & \frac{1}{f_\star^2} \left\{ 1 - 2 \int_0^\infty
\frac{du}{u} \, w_\theta(x_\star,u) \, \frac{f'}{f} + 2 \left[
\int_0^\infty \frac{du}{u} \, w_\theta(x_\star,u) \, \frac{f'}{f}
\right]^2 + 2 \left[ \int_0^\infty \frac{du}{u} \, \chi(u) \,
\frac{f'}{f} \right]^2 \right. \nonumber
\\ & &
\left. \hspace{2em} \mbox{} - 4 \int_0^\infty \frac{du}{u} \,
\chi(u) \, \frac{f'}{f} \int_u^\infty \frac{dv}{v^2} \,
\frac{f'}{f} + \mathcal{O} \left[ \left( \frac{f'}{f} \right)^3
\right] \right\}.
\end{eqnarray}
In addition to the general slow-roll assumptions, standard
slow-roll assumes
\begin{equation}\label{src}
\frac{1}{f} \frac{d^n f}{(d\ln x)^n} =
\mathcal{O}\left[ \left( \frac{f'}{f}
\right)^n \right]
\end{equation}
in which case we can Taylor expand $f$ in terms of $\ln(x/x_*)$
where $x_\star$ is some convenient time around horizon crossing
\begin{eqnarray}
\frac{f'}{f} & = & \frac{f'_\star}{f_\star} + \left( \frac{f'}{f}
\right)'_\star \ln \frac{x}{x_\star} + \frac{1}{2} \left(
\frac{f'}{f} \right)''_\star \left( \ln \frac{x}{x_\star}
\right)^2 + \cdots \nonumber
\\
& = & \frac{f'_\star}{f_\star} + \left[ \frac{f''_\star}{f_\star}
- \left( \frac{f'_\star}{f_\star} \right)^2 \right] \ln
\frac{x}{x_\star} + \frac{1}{2} \left[ \frac{f'''_\star}{f_\star}
- 3 \frac{f'_\star}{f} \frac{f''_\star}{f} + 2 \left(
\frac{f'_\star}{f_\star} \right)^3 \right] \left( \ln
\frac{x}{x_\star} \right)^2 + \cdots.
\nonumber \\
\end{eqnarray}
Substituting into Eq.~(\ref{PSf}) gives
\begin{eqnarray}\label{pf}
\mathcal{P} & = & \frac{1}{f_\star^2} \left\{ 1 - 2\alpha_\star
\frac{f'_\star}{f_\star} + \left( -\alpha_\star^2 +
\frac{\pi^2}{12} \right) \frac{f''_\star}{f_\star} + \left(
3\alpha_\star^2 - 4 + \frac{5\pi^2}{12} \right) \left(
\frac{f'_\star}{f_\star} \right)^2 \right. \nonumber
\\ & &
\hspace{3em} \mbox{} + \left[ - \frac{1}{3}\alpha_\star^3 +
\frac{\pi^2}{12}\alpha_\star - \frac{4}{3} + \frac{2}{3}\zeta(3)
\right] \frac{f_\star'''}{f_\star} \nonumber
\\ & &
\hspace{3em} \mbox{} + \left[ 3\alpha_\star^3 - 8\alpha_\star +
\frac{7}{12}\pi^2\alpha_\star + 4 - 2\zeta(3) \right]
\frac{f_\star'f_\star''}{f_\star^2} \nonumber
\\ & &
\left. \hspace{3em} \mbox{} + A \left( \frac{f_\star'}{f_\star}
\right)^3 + \mathcal{O} \left[ \left( \frac{f_\star'}{f_\star}
\right)^4 \right] \right\}
\end{eqnarray}
where
\begin{equation}
\alpha_\star \equiv \alpha - \ln x_\star,
\end{equation}
\begin{equation}
\alpha \equiv 2 - \ln 2 - \gamma \simeq 0.729637,
\end{equation}
$\gamma \simeq 0.577216$ is the Euler-Mascheroni constant, $\zeta$
is the Riemann zeta function, and $A$ is an undetermined
coefficient. This is consistent with the second order standard
slow-roll result, Eq.~(40) in Ref.~\cite{ps2nds}, where the result
was evaluated at $x_\star = 1$. However, because $A$ is the only
undetermined third order coefficient, we can use the known exact
solutions \cite{ps1st} to determine the complete third order
standard slow-roll result.\footnote{It was brought to our
attention that the authors of Ref.~\cite{3rdex} used these exact
solutions to determine the third order corrections valid under
some special conditions. Note that we use them to help determine
the completely general case.} In fact, if we were sufficiently
motivated, we could apply the same method to determine the fourth
order standard slow-roll result from Eq.~(\ref{PSf}).

The simplest exact solution is inflation near a maximum, where the
potential is
\begin{equation}
V(\phi) = V_0 \left( 1 - \frac{1}{2} \mu^2 \phi^2 + \cdots
\right).
\end{equation}
In this case
\begin{equation}
H = \sqrt{\frac{V_0}{3}}\,, \ \ \ x = \frac{k}{aH}
\end{equation}
and
\begin{equation}
f = \frac{3\pi \phi_0}{H^2} \left( \sqrt{1+\frac{4}{3}\mu^2} - 1
\right) \left( \frac{x}{x_0} \right)^{- \frac{3}{2} \left(
\sqrt{1+\frac{4}{3}\mu^2} - 1 \right)}.
\end{equation}
Then, by comparing the exact solution \cite{ps1st} and
Eq.~(\ref{pf}), we obtain
\begin{equation}\label{A}
A = -4\alpha_\star^3 + 16\alpha_\star -
\frac{5}{3}\pi^2\alpha_\star - 8 + 6\zeta(3).
\end{equation}
We can check this result by using the exact solution for power law
inflation, where
\begin{equation}
V(\phi) = V_0 \exp \left( - \sqrt{\frac{2}{p}} \, \phi \right), \
\ \ x = \left(\frac{p}{p-1}\right) \left(\frac{k}{aH}\right)
\end{equation}
and
\begin{equation}
f = \frac{2\pi\sqrt{2p}}{(p-1)H_0} \left( \frac{x}{x_0}
\right)^{-\frac{1}{p-1}}
\end{equation}

The slow-roll parameters are defined as
\begin{equation}
\epsilon = - \frac{\dot H}{H^2} = \frac{1}{2}
\left(\frac{\dot{\phi}}{H}\right)^2 \ \ \mbox{and} \ \ \delta_n =
\frac{1}{H^n \dot\phi} \frac{d^n \dot\phi}{dt^n} \, ,
\end{equation}
where in standard slow-roll
\begin{equation}
\epsilon = \mathcal{O}(\xi)
\ \ \mbox{and} \ \
\delta_n = \mathcal{O}(\xi^n)
\end{equation}
for some small parameter $\xi$.
Then
\begin{eqnarray}
\frac{1}{f^2} & = & \left( \frac{H}{2\pi} \right)^2 \left(
\frac{H}{\dot\phi} \right)^2 \left[ 1 - 2\epsilon - 3\epsilon^2 -
4\epsilon\delta_1 - 4\epsilon\delta_2 - 16\epsilon^3 -
28\epsilon^2\delta_1 - 4\epsilon\delta_1^2 + \mathcal{O}(\xi^4)
\right], \nonumber
\\
\frac{f'}{f} & = & -2\epsilon - \delta_1 - 4\epsilon^2 -
3\epsilon\delta_1 - 2\epsilon\delta_2 - 18\epsilon^3 -
25\epsilon^2\delta_1 - 4\epsilon\delta_1^2 + \mathcal{O}(\xi^4),
\nonumber
\\
\frac{f''}{f} & = & \delta_2 + 8\epsilon^2 + 9\epsilon\delta_1 +
4\epsilon\delta_2 + 36\epsilon^3 + 50\epsilon^2\delta_1 +
8\epsilon\delta_1^2 + \mathcal{O}(\xi^4), \nonumber
\\
\frac{f'''}{f} & = & - \delta_3 - 13\epsilon\delta_2 -
48\epsilon^3 - 85\epsilon^2\delta_1 - 18\epsilon\delta_1^2 +
\mathcal{O}(\xi^4),
\end{eqnarray}
where now the standard slow-roll conditions, Eq.~(\ref{src}), are
manifest in terms of the slow-roll parameters. Substituting into
Eq.~(\ref{pf}) and using Eq.~(\ref{A}), we obtain
\begin{eqnarray}\label{PSsr-star}
\mathcal{P} & = & \left( \frac{H_\star}{2\pi} \right)^2 \left(
\frac{H_\star}{\dot\phi_\star} \right)^2 \left\{ 1 + (
4\alpha_\star - 2 ) \epsilon_\star + 2\alpha_\star \delta_{1\star}
+ \left( -\alpha_\star^2 + \frac{\pi^2}{12} \right)
\delta_{2\star} \right. \nonumber
\\ & &
\mbox{} + \left( 4\alpha_\star^2 - 19 + \frac{7\pi^2}{3} \right)
\epsilon_\star^2 + \left( 3 \alpha_\star^2 + 2\alpha_\star - 20 +
\frac{29\pi^2}{12} \right) \epsilon_\star\delta_{1\star} + \left(
3\alpha_\star^2 - 4 + \frac{5\pi^2}{12} \right) \delta_{1\star}^2
\nonumber
\\ & &
\mbox{} + \left[ \frac{1}{3}\alpha_\star^3 - \frac{\pi^2}{12}
\alpha_\star + \frac{4}{3} - \frac{2}{3}\zeta(3) \right]
\delta_{3\star} \nonumber
\\ & &
\mbox{} + \left[ -\frac{5}{3}\alpha_\star^3 - 2\alpha_\star^2 +
20\alpha_\star - \frac{9}{4}\pi^2\alpha_\star + \frac{16}{3} +
\frac{\pi^2}{6} - \frac{14}{3}\zeta(3) \right]
\epsilon_\star\delta_{2\star} \nonumber
\\ & &
\mbox{} + \left[ -3\alpha_\star^3 + 8\alpha_\star -
\frac{7}{12}\pi^2\alpha_\star - 4 + 2\zeta(3) \right]
\delta_{1\star}\delta_{2\star} \nonumber
\\ & &
\mbox{} + \left[ 4\alpha_\star^2 + 8\alpha_\star + 16 + 5\pi^2 -
48\zeta(3) \right] \epsilon_\star^3 \nonumber
\\ & &
\mbox{} + \left[ -\frac{5}{3}\alpha_\star^3 + 4\alpha_\star^2 +
32\alpha_\star - \frac{9}{4}\pi^2\alpha_\star + \frac{88}{3} +
\frac{23}{3}\pi^2 - \frac{230}{3}\zeta(3) \right]
\epsilon_\star^2\delta_{1\star} \nonumber
\\ & &
\mbox{} + \left[ 3\alpha_\star^3 + 4\alpha_\star^2 -
24\alpha_\star + \frac{13}{4}\pi^2\alpha_\star + 16 +
\frac{7}{3}\pi^2 - 30\zeta(3) \right]
\epsilon_\star\delta_{1\star}^2 \nonumber
\\ & &
\left. \mbox{} + \left[ 4\alpha_\star^3 - 16\alpha_\star +
\frac{5}{3}\pi^2\alpha_\star + 8 - 6\zeta(3) \right]
\delta_{1\star}^3 \right\}.
\end{eqnarray}
We see that this third order result is consistent with the second
order result of Eq.~(41) of Ref.~\cite{ps2nds} if we note that the
result of Ref.~\cite{ps2nds} was evaluated at $a_\star H_\star =
k$, in which case we have $\alpha_\star \equiv \alpha - \ln
x_\star = \alpha - \epsilon_\star + \mathcal{O}(\xi^2)$.

\subsection*{Acknowledgements}

It is a great pleasure to thank Seung-Ho Lee for collaboration
during the early stages of this work, and we are indebted to
Hyun-Chul Lee for invaluable discussions at every stage of this
work. We are also grateful to Misao Sasaki and Takahiro Tanaka for
helpful discussions and suggestions, and the Yukawa Institute for
Theoretical Physics for hospitality while this work was in
progress. We thank Jai-chan Hwang for helpful comments on a draft
of this paper. This work was supported in part by ARCSEC funded by
the Korea Science and Engineering Foundation and the Korean
Ministry of Science, the Korea Research Foundation grant KRF PBRG
2002-070-C00022, and Brain Korea 21.

\end{document}